\documentclass[prb,aps,twocolumn,amsmath,amssymb]{revtex4}
\usepackage{graphicx}
\usepackage{dcolumn}
\usepackage{bm}
\usepackage{times}
\usepackage{epstopdf}
\usepackage{amsmath}
\usepackage{epsfig}
\usepackage{comment}

\begin{document}
\title{Partial positive refraction in asymmetric Veselago lenses of uniaxially strained graphene}

\author{Y. Betancur-Ocampo}
\email{ybetancur@icf.unam.mx}
\affiliation{Instituto de Ciencias F\'isicas, Universidad Nacional Aut\'onoma de M\'exico, Cuernavaca, M\'exico}
\date{\today}

\begin{abstract}

Asymmetric Veselago lenses (AVLs) can be created from ballistic $p$-$n$ and $n$-$p$-$n$ homojunctions of uniaxially strained graphene. This atypical converging electron flow emerges by applying uniaxial tension out of the device's symmetry axes. A part of electron flow needs to be positively refracted for focusing in an asymmetric spot, whose location is tunable with the strain. In AVLs, Klein tunneling is angularly shifted regards to the normal incidence. This perfect transmission occurs at the straight line that connects the point source and focus, which is unaffected by variation of the Fermi level and barrier's width. Moreover, the mirror symmetry breaking by the strain also causes the asymmetry in Fabry-P\'erot interference. The novel electron optical laws allow to evidence that reflected and refracted electrons in AVLs lie on the same straight line with opposite group velocities and pseudo-spins. Unlike isotropic graphene, electrons under normal incidence present backscattering, angles of reflection and refraction different to zero. The average particle transmission is higher (lower) than isotropic case when the tensile strain is increased near (far away) the normal direction. These results may be useful for designing strain-bendable probing tips in scanning tunneling microscopes.
\end{abstract}


\maketitle

\section{Introduction}

Strain-engineering in graphene has opened a wide range of possibilities to control the electronic and transport properties \cite{Naumis,Naumis2,Betancur,Stegmann,Pellegrino,Pellegrino2,Pereira,Pereira2,Pereira3,Ribeiro,Choi}. Initially, it was proposed to generate a gap opening in graphene without detriment of its outstandings properties. Thus, the strain-induced energy band gap could be performed through a topological phase transition from the semimetallic to insulator phase \cite{Pereira}. Currently, the study of strain effects on the electronic band structure in graphene is considered in order to modulate physical properties \cite{Pereira,Pellegrino,Betancur}. Thus, the generation of pseudo-magnetic fields \cite{Guinea,Levy,Gomes,Vozmediano} and the development of valleytronics have been addressed \cite{Ang,Glattli,Zhai,Jiang,Peeters,Zhai2,Charlier,Li,Pomar}. On the other hand, electron optics in graphene offers the opportunity of manipulating the trajectories of electrons for concrete applications in nanoelectronics \cite{Katsnelson,Kim,Khan,Beenakker,Ojanen,Betancur2,Betancur3,Fuchs,Gunlycke,Ho Lee,Chen,Rickhaus,Williams,Liu,Park,Louie,Richter,Masum,Boggild,Hills,Cheianov}. Recently, negative refraction of massless Dirac fermions was tested \cite{Ho Lee,Chen}. This observation paves the way for the use of Veselago lenses to control the electron flow such as the light in metamaterials \cite{Cheianov,Veselago,Pendry,Schurig}. The merging of strain-engineering and electron optics may provide a broad perspective to investigate novel and unusual phenomena in graphene and other Dirac materials.

Many contributions in strained graphene are dedicated to the control of valley spin polarization, where the main motivation is to use the valley degree of freedom as conveyor of quantum information \cite{Ang,Glattli}. This proposal arises due to the valley-dependence of refraction when electrons tunnel from unstrained to strained graphene regions \cite{Zhai,Zhai2,Jiang,Peeters,Charlier}. However, obtaining a well-defined isotropic and anisotropic graphene regions must be difficult in the practice. While $p$-$n$ and $n$-$p$-$n$ homojunctions of uniaxially strained graphene (USG) is more factible to test transport phenomena, it has been scarcely adressed due to the valley-independence in the electron transport. Nevertheless, unusual optical-like phenomena emerges considering fully strained graphene sheets. In this work is shown that in-plane deformation out of the symmetry axes of a graphene $p$-$n$ junction creates asymmetric Veselago lenses (AVLs). To difference of those perfect lenses in metamaterials and isotropic graphene which focus the particle flow towards a symmetric spot \cite{Cheianov}, strained graphene homojunctions can bend the converged electron flow. Thus, a part of electron flow is positively refracted for focusing in an asymmetric spot. In these systems, reflected and refracted electrons have inverted pseudo-spins and move in opposite directions. Further, the Klein tunneling (KT) is angularly shifted and occurs at the straight line that links the point source and focus. The particle transmission efficiency in AVLs is higher (lower) than unstrained case when the tensile strain is increased near (far away) the normal direction. Since the Fabry-P\'erot interference of $n$-$p$-$n$ homojunctions has not mirror symmetry, resonant tunneling under normal incidence appears. Such results may be useful for designing an improved scanning tunneling microscope with bendable probing tip. 

This paper is organized as follow: in the section \ref{model} is used the Tight-Binding (TB) approach to nearest neighbors for anisotropic graphene sheet. In this calculation the uniaxial strain modifies the electronic band structure changing the rotation and shape of Dirac cones. An effective TB hamiltonian is obtained through Taylor expansion around the Dirac points up to first order in wave vector. Thus, the complex velocity parameters of Weyl-like Hamiltonian are expressed in terms of uniaxial strain components. In the section \ref{opt_laws} is established the reflection and refraction laws of electrons impinging the interface of homojunctions. The application of these novel electron optics laws in the section \ref{AVLs} evidences how partial positive refraction of a particle flow favors to the creation of AVLs. In this same section is shown the angular deviation of KT and average particle transmission efficiency. Whereas the shifting of Fabry-P\'erot fringes of USG $n$-$p$-$n$ homojunctions are discussed in the section \ref{npn_junc}. Conclusions and final remarks are exposed in the section \ref{conclusions}. 

\section{Complex velocities and effective TB Weyl-like Hamiltonian of uniaxially strained graphene}
\label{model} 

\begin{figure}
\begin{tabular}{c}
(a) \qquad \qquad \qquad \qquad \qquad \qquad \qquad \qquad \qquad \qquad \qquad \qquad \qquad\\
\includegraphics[trim = 0mm 0mm 0mm 0mm, scale= 0.20, clip]{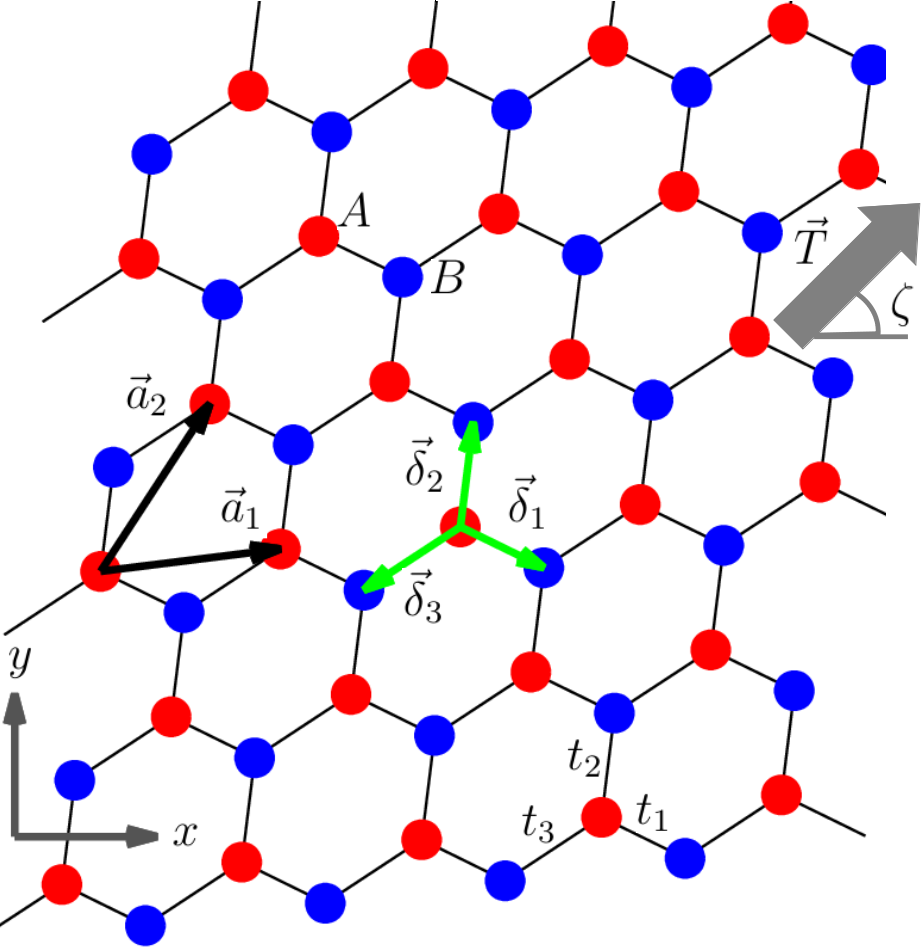}\\
(b) \qquad \qquad \qquad \qquad \qquad \qquad \qquad \qquad \qquad \qquad \qquad \qquad \qquad\\
\includegraphics[trim = 0mm 0mm 0mm 0mm, scale= 0.20, clip]{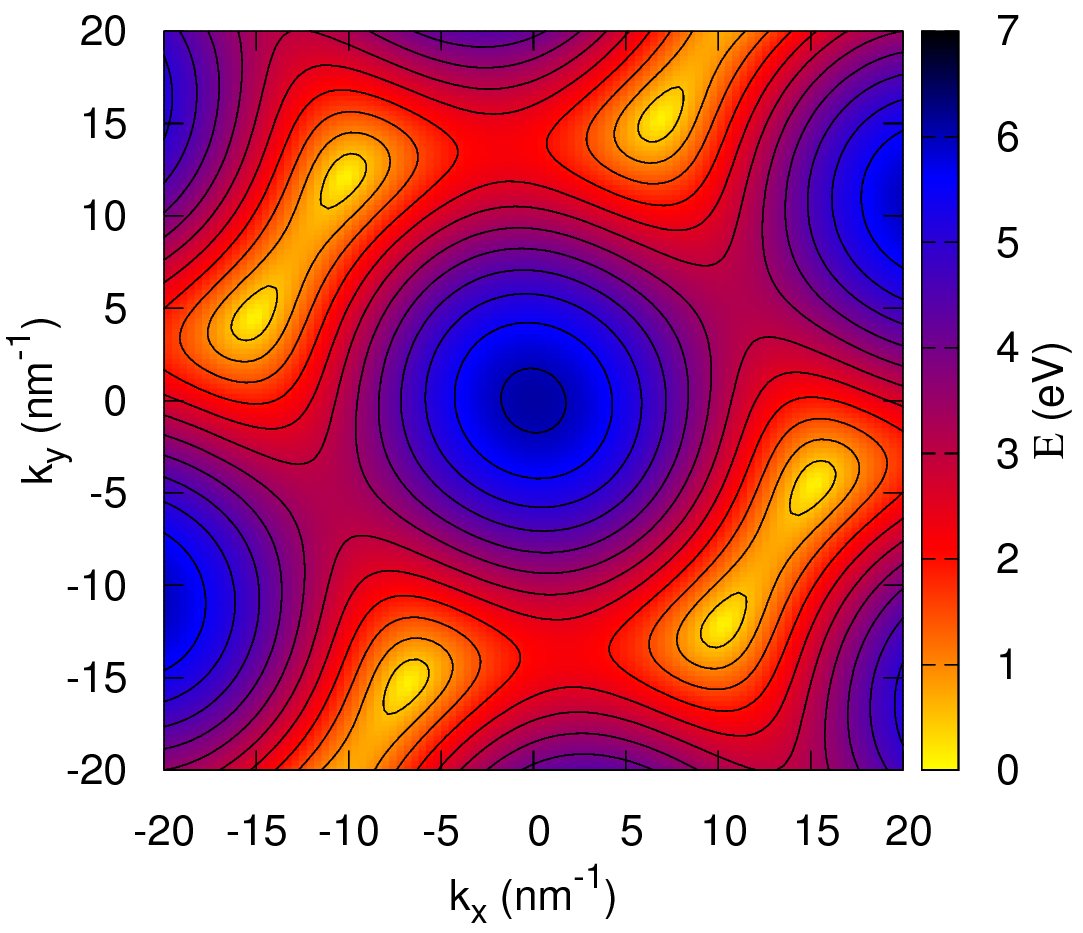}
\end{tabular}
\caption{(a) Schematic representation of uniaxially strained graphene. Red and blue circles correspond to the sites of triangular sublattices $A$ and $B$, respectively. The $x$ axis matches with the zigzag bond in the unstrained case. The uniaxial tension $\vec{T}$ is applied along the $\zeta$ direction. The strained configuration has three different hopping parameters $t_j$ and bond lengths $\delta_j$. The lattice vectors are denoted by $\vec{a}_1$ and $\vec{a}_2$. (b) Contour energy density near to the first Brillouin zone for the conduction band obtained from the TB approach to nearest neighbors using the set of values $\epsilon = 23 \%$ and $\zeta = 45^{\textrm{o}}$.}
\label{system}
\end{figure}

USG is constituted by two deformed triangular Bravais sublattices which are labeled as $A$ and $B$, [see Fig. \ref{system}(a)]. This deformed crystal possesses a unit cell having two carbon atoms with decoupled $p_z$ and $\sigma$ orbitals. The atomic sites are displaced applying the tension $T$ in the direction $\zeta$. The Cartesian system is set with the $x$ axis along the $ZZ$ bond in the unstrained configuration \cite{Pereira}. The positions of nearest neighbors are denoted by $\vec{\delta_1}$, $\vec{\delta_2}$ and $\vec{\delta_3}$ on the underlying sublattice $A$. Whereas the lattice vectors $\vec{a}_1$ and $\vec{a}_2$ and reciprocal ones $\vec{b}_1 = (2\pi/|\vec{a}_1\times\vec{a}_2|)\vec{a}_2\times\hat{z}$ and $\vec{b}_2 = (2\pi/|\vec{a}_1\times\vec{a}_2|)\hat{z}\times\vec{a}_1$ build the whole positions in the deformed hexagonal lattices, where $\hat{x}$, $\hat{y}$, and $\hat{z}$ are unit vectors of the Cartesian system. The first Brillouin zone corresponds to a distorted hexagon with two nonequivalent high symmetry points $K$ and $K'$. From elasticity theory \cite{Pereira,Landau,Colombo,Cadelano}, the uniaxial strain tensor is written as

\begin{equation}
{\bf u} = (p{\bf I} + q{\bf S})\epsilon,
\label{str}
\end{equation}

\noindent where the constants $p$ and $q$ are defined by $p = (1 - \nu)/2$ and $q = (1 + \nu)/2$, being $\nu$ the Poisson ratio of graphene \cite{Cadelano,Betancur,Ribeiro,Castro,Pereira}. The identity matrix ${\bf I}$ has dimension $2\times2$ as well as ${\bf S}$, which is expressed in terms of the Pauli matrices ${\bf S} = {\bf \sigma}_z\cos2\zeta + {\bf \sigma}_x\sin2\zeta$. The tensile strain $\epsilon$ quantifies the tension magnitude and the tension angle $\zeta$ indicates its direction, as shown in Fig. \ref{system}(a). Since ${\bf u}$ is a homogeneous strain tensor, the displacement vector is $\vec{r} = ({\bf I} + {\bf u})\vec{r}_o$ being $\vec{r}_o$ the vector on the old configuration. Hence, the deformed lattice vectors  

\begin{eqnarray}
\vec{a}_1 & = & \sqrt{3}a[\hat{x}(1 + p\epsilon + q\epsilon\cos2\zeta) + \hat{y}q\epsilon\sin2\zeta] \nonumber\\
\vec{a}_2 & = & \frac{\sqrt{3}}{2}a\{\hat{x}[1 + p\epsilon + 2q\epsilon\cos(2\zeta -\pi/3)] + \nonumber\\
 & & \hat{y}[\sqrt{3}(1 + p\epsilon) + 2q\epsilon\sin(2\zeta -\pi/3)]\},
\label{lattvs}
\end{eqnarray}

\noindent are related with the uniaxial strain parameters, where $a$ is the bond length in pristine graphene \cite{Castro}. Thus, the relative nearest neighbors sites for USG $\vec{\delta}_1 = 2\vec{a}_1/3 - \vec{a}_2/3$, $\vec{\delta}_2 = 2\vec{a}_2/3 - \vec{a}_1/3$, and $\vec{\delta}_3 = -\vec{\delta}_1 -\vec{\delta}_2$ are also obtained.

In order to calculate the energy band structure from the TB approach is needed to establish the relation of hopping parameters with uniaxial strain tensor. These TB parameters are the probability amplitudes that an electron in sublattice $A$ hops to neighboring sites. The hoppings can be modeled with an exponential decay rule $t_j = t\textrm{exp}[-\beta(\delta_j/a - 1)]$, where $\beta$ is the Gr\"uneisen constant, $t$ is the hopping in pristine graphene, and $\delta_j$ are the deformed bond lengths \cite{Ribeiro,Pereira,Papas,Castro,Betancur}. The complete relation of $t_j$ as a function of strain parameters is given by 

\begin{widetext}
\begin{equation}
\delta_j = a\sqrt{\left[1 + p\epsilon + q\epsilon\cos{\left(2\zeta + \tfrac{2j - 1}{3}\pi\right)}\right]^2 + \epsilon^2q^2\sin^2\left(2\zeta + \tfrac{2j - 1}{3}\pi\right)}.
\label{d}
\end{equation}
\end{widetext}

\noindent Since the overlap terms and the next nearest neighbors hopping have a negligible contribution in graphene \cite{Goerbig}, TB Hamiltonian is only considered up to nearest neighbors

\begin{equation}
H_{TB} = \sum^3_{j = 1}\left[\begin{array}{cc}
0 & t_j\textrm{e}^{i\vec{k}\cdot\vec{\delta}_j}\\
t_j\textrm{e}^{-i\vec{k}\cdot\vec{\delta}_j} & 0
\end{array}\right].
\label{H}
\end{equation} 

\noindent Therefore, electronic band structure of USG is obtained from the eigenvalues of the TB Hamiltonian \eqref{H}  

\begin{equation}
E_{\pm} = s\sqrt{\sum^3_{i = 1}t^2_i + 2\sum^3_{i<j}t_it_j\cos[\vec{k}\cdot(\vec{\delta}_i - \vec{\delta}_j)]},
\label{dr}
\end{equation}

\noindent where the band index $s = \textrm{sgn}(E)$ indicates the valence ($s = -1$) and conduction ($s = 1$) energy band. The corresponding eigenstates of the Hamiltonian \eqref{H} are expressed as $| \Psi(\vec{k})\rangle = \tfrac{1}{\sqrt{2}}(1,s\textrm{e}^{i\phi(\vec{k})})$, where

\begin{equation}
\phi = -\arctan\left(\frac{\sum_jt_j\sin(\vec{k}\cdot\vec{\delta}_j)}{\sum_jt_j\cos(\vec{k}\cdot\vec{\delta}_j)}\right)
\end{equation}

\noindent is the pseudo-spin angle. The anisotropy-induced by the application of uniaxial strain causes the distortion of energy bands, as shown in Fig. \ref{system}(b). Contour curves around Dirac points evolve from elliptical to nontrivial shape when the energy is increased. If TB Hamiltonian \eqref{H} is expanded around the Dirac point performing $\vec{k} = \vec{q} + \vec{K}_D$, where the Dirac point position $\vec{K}_D$ satisfies $\sum_j\textrm{exp}(i\vec{K}_D\cdot\vec{\delta}_j) = 0$, the effective Weyl-like Hamiltonian

\begin{equation}
H_W = \left[\begin{array}{cc}
0 & w^{c*}_xp_x + w^{c*}_yp_y\\
w^c_xp_x + w^c_yp_y & 0
\end{array}\right]
\label{HD}
\end{equation}

\noindent is expressed in terms of complex velocities $\vec{w}_c$. These velocities are defined as 
\begin{equation}
 \vec{w}_c = (w^c_x,w^c_y) =i\sum^3_{j = 1}\frac{\vec{\delta}_j}{\hbar}t_j\textrm{e}^{-i\vec{K}_D\cdot\vec{\delta}_j},
 \label{w}
\end{equation} 

\noindent which establish a relation with the strain parameters through the lattice quantities and electronic contributions. The Hamiltonian \eqref{HD} can be led to the standard form $H_W = v_{ij}\sigma_ip_j$, if the complex velocities components $w^c_x$ and $w^c_y$ are written as $w^c_x = w_x\textrm{e}^{-i\alpha_x}$ and $w^c_y = w_y\textrm{e}^{i\alpha_y}$, where $\alpha_x$ and $\alpha_y$ are the velocity phases. Useful identities of complex velocities with hopping parameters and lattice vectors are exposed in appendix A. On the other hand, a similar form of the Hamiltonian \eqref{HD} is obtained from the other non-equivalent Dirac point $-\vec{K}_D$. Thus, it is possible to construct a 4 $\times$ 4 block-diagonal matrix representation $H = \tau_z\otimes H_W$, where $\tau_z$ is the $z$-component Pauli matrix acting on pseudospin space and discriminating the contribution of the pseudospin valleys \cite{Goerbig}. 

\section{Optical laws of massless Dirac fermions in anisotropic media}  
\label{opt_laws}

\begin{figure}
\begin{tabular}{c}
(a) \qquad \qquad \qquad \qquad \qquad \qquad \qquad \qquad \qquad \qquad \qquad \qquad \qquad\\
\includegraphics[trim = 0mm 0mm 0mm 0mm, scale= 0.33, clip]{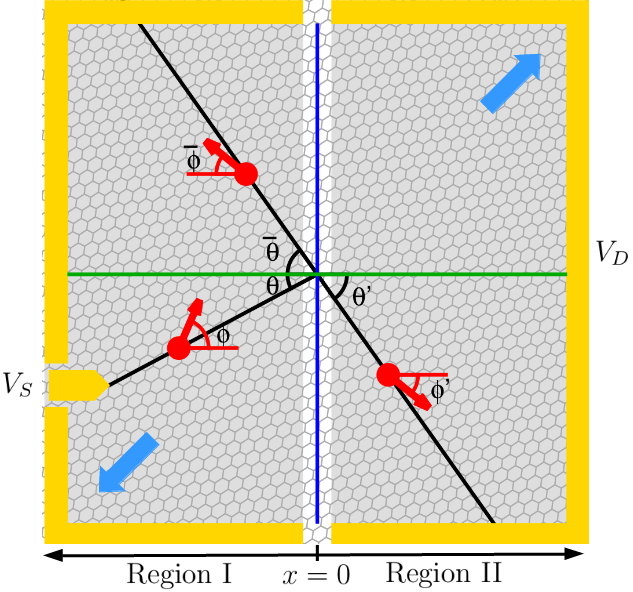}\\
(b) \qquad \qquad \qquad \qquad \qquad \qquad \qquad \qquad \qquad \qquad \qquad \qquad \qquad\\
\includegraphics[trim = 0mm 0mm 0mm 0mm, scale= 0.28, clip]{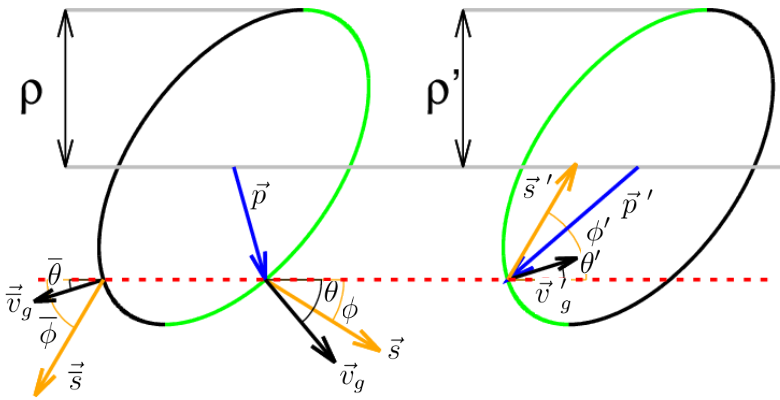}
\end{tabular}
\caption{Geometrical optics convention of angles and kinematical construction in $p$-$n$ homojunction of uniaxially strained graphene. (a) Incoming electron beam with angle of incidence $\theta$ and pseudo-spin angle $\phi$ is scattered at the interface. Outgoing electron rays with angles of pseudo-spin $\overline{\phi}$ ($\phi'$) and reflection $\overline{\theta}$ (refraction $\theta'$) obey atypical reflection (refraction) laws \eqref{phirefl} and \eqref{eoptlr} [\eqref{phi_Snell} and \eqref{eoptlrr}], respectively. (b)  Kinematical construction shows the scattering of electrons in the reciprocal space. The ellipses correspond to the energy contour at the Fermi level $E = V_0/2$ for both regions. The dashed red line (green semiarcs) represents the conservation of linear momentum $p_y$ (probability current density $j_x$). Refraction index $\rho$ ($\rho'$) is the vertical half-width of ellipse in the region I (II). The black and golden arrows denote the direction of group velocity and pseudo-spin angles, respectively.}
\label{EOD}
\end{figure}

Optical-like phenomena can be simulated from ballistic $p$-$n$ homojunctions of USG, as shown in Fig. \ref{EOD}(a). The external split-gate structure $V$ and $V'$ creates an abrupt step potential between the region I and II which warranties the transmission by propagation modes. The necessary conditions for obtaining electron optics require to lead the system to a ballistic regime \cite{Low2,Cheianov2,Fuchs}. Such special conditions have been recently achieved in the experimental observation of negative refraction of electrons \cite{Ho Lee,Chen}. In this device, the linear interface separates two regions with different charge density (see Fig. \ref{EOD}). Thus, electrons tunnel on the interface changing their group velocity and pseudo-spin. This is essentially important for controlling electron flow through the tuning of strain parameters and Fermi level. The point source $V_S$ in Fig. \ref{EOD}(a) spreads electrons in the whole directions. In order to avoid unwanted interference by multiple reflections at the borders, an extended bias $V_D$ voltage drains the output electrons. When particles impinges on the interface, the optical laws describe the redirection of reflected and refracted flow as well as the scattering probability. The conservation of energy $E$, linear momentum $p_y$, and probability current density $j_x$ are useful for establishing these electron optics laws in USG, which are schematically represented by the kinematical construction in Fig. \ref{EOD}(b). 

\begin{figure*}
\begin{tabular}{cc}
(a) \qquad \qquad \qquad \qquad \qquad \qquad \qquad \qquad \qquad \qquad \qquad & (b) \qquad \qquad \qquad \qquad \qquad \qquad \qquad \qquad \qquad \qquad \qquad \\
\includegraphics[trim = 0mm 0mm 0mm 0mm, scale= 0.15, clip]{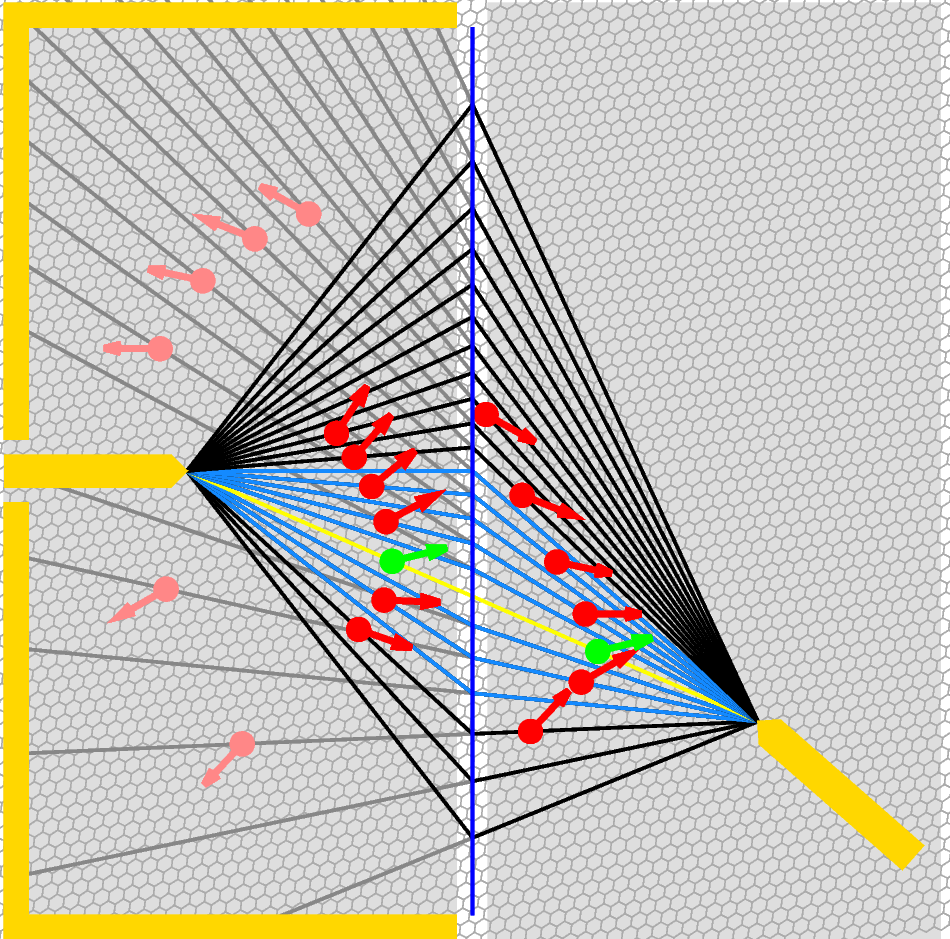} &
\includegraphics[trim = 0mm 0mm 0mm 0mm, scale= 0.3, clip]{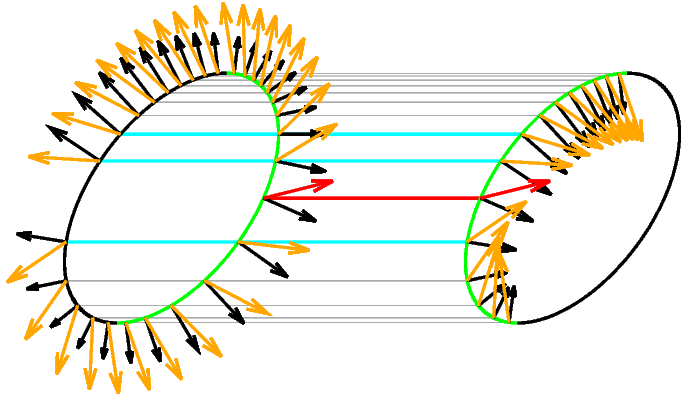}\\
(c) \qquad \qquad \qquad \qquad \qquad \qquad \qquad \qquad \qquad \qquad \qquad & (d) \qquad \qquad \qquad \qquad \qquad \qquad \qquad \qquad \qquad \qquad \qquad \\
\includegraphics[trim = 0mm 0mm 0mm 0mm, scale= 0.2, clip]{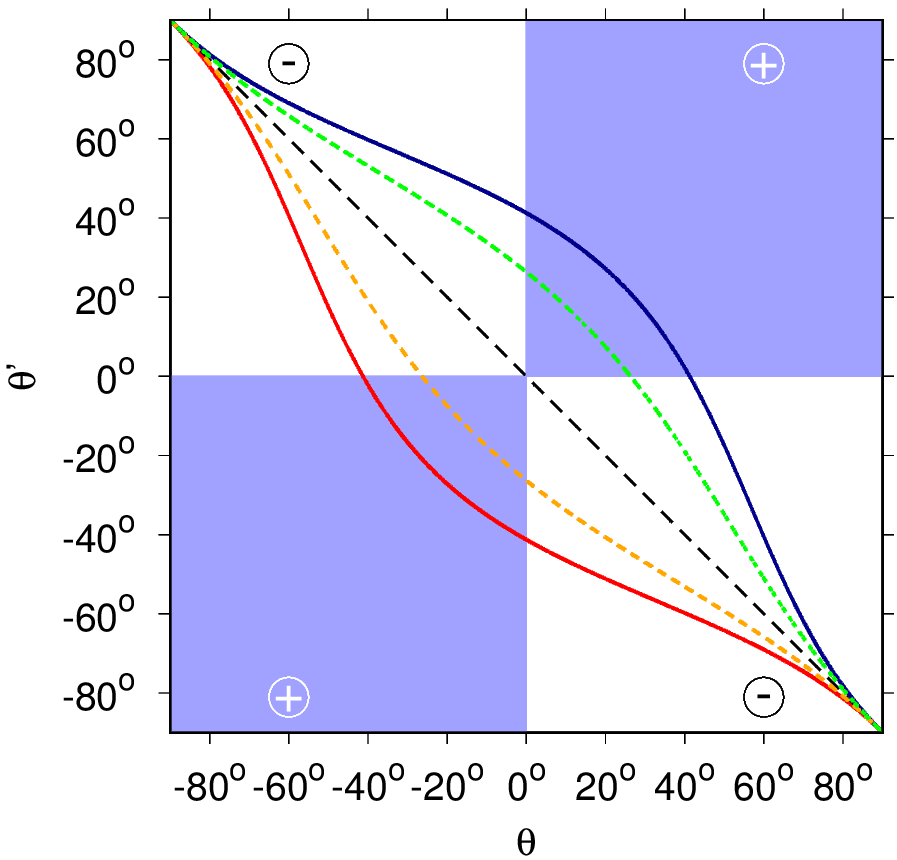} &
\includegraphics[trim = 0mm 0mm 0mm 0mm, scale= 0.2, clip]{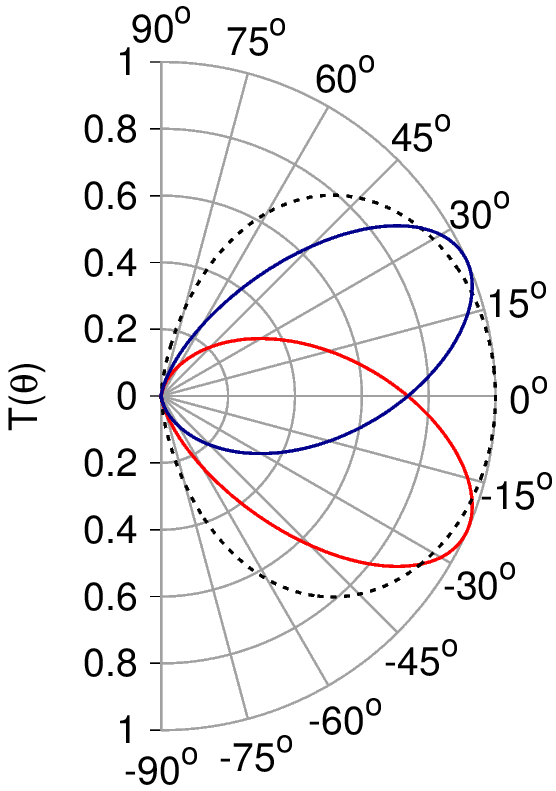}
\end{tabular}
\caption{Partial positive refraction and angular shifting of Klein tunneling in an asymmetric Veselago lens. (a) The special lens can be obtained applying the tensile strain $\epsilon = 23 \%$ along the direction $\zeta = 45^{\textrm{o}}$. When the point source at $(-x_0,0)$ spreads electrons towards the interface, the refracted flow meets at the spot $(x_0,y'_0)$ for $E = 50$ meV. The blue rays indicate the beams with positive refraction. The yellow straight line, which links the point source and focus, is the Klein tunneling path with deviation angle $\theta_{KT} = -23.7^{\textrm{o}}$. (b) Kinematical construction reveals how the positive refraction, atypical reflection, and deviation of Klein tunneling emerge. Black (red and golden) arrows correspond to group velocities (pseudo-spins). Whereas horizontal lines denote the conservation of $p_y$. The pseudo-spin and group velocity are conserved for $p_y = 0$ (red line). The turquoise horizontal lines show the positive refraction of group velocity. (c) Angle of refraction $\theta'$ as a function of $\theta$ for the set of values $\epsilon = 23 \%$ and $\zeta = 45^{\textrm{o}}$ (red curve), $15^{\textrm{o}}$ (dashed orange), $0^{\textrm{o}}$ (dashed black), $-15^{\textrm{o}}$ (dashed green), and $-45^{\textrm{o}}$ (blue). White and blue regions point up negative and positive refraction, respectively. (d) Transmission probability as a function of $\theta$ for the same set of parameters in (c).}
\label{CAVLs}
\end{figure*}

The refraction index $\rho$ has a direct geometrical meaning and corresponds to the vertical half-width of the elliptical energy contour at the Fermi level [see Fig. \ref{EOD}(b)]. The primed quantities are denoted for the region II. It is possible to establish an auxiliar Snell's law through the conservation of $p_y$ given by $\rho\sin\chi = \rho'\sin\chi'$, where $\chi$ and $\chi'$ are effective angles. Although these quantities differ of the genuine angles of scattering $\theta$ and $\theta'$, which are given by the group velocity direction, the angles $\chi$ are useful for symplifying the optical laws. Using the dispersion relation of Hamiltonian \eqref{HD}, the components of linear momentum can be expressed as $p_x = \pm s(w_y/w_x)\rho\sin(\alpha_x + \alpha_y \mp \chi)$ and $p_y = \rho\sin\chi$ (see appendix B). The minus in $p_x$ indicates reflection for the region I. From geometrical arguments, it is shown that the refraction index is $\rho = |E - V|/[w_y\sin(\alpha_x + \alpha_y)]$. Thus, the definition of pseudo-spin angle and conservation of $p_y$ allow to derive the following relations (see appendix B) 

\begin{equation}
\phi = s\chi + \alpha_x, \quad \overline{\phi} = s\chi -\alpha_x, \quad \textrm{and} \quad \phi' = s'\chi' + \alpha_x,
\label{phis}
\end{equation}

\noindent where the geometrical optics convention for angles is used. The overline denotes the reflected quantities. In a standard form, the reflection law of pseudo-spin for anisotropic massless Dirac fermions is written as

\begin{equation}
\overline{\phi} = \phi - 2\alpha_x.
\label{phirefl}
\end{equation}

\noindent It is worth to note that reflected electron beams do not obey the conventional reflection law $\phi = \overline{\phi}$. This is due to that uniaxial strain along the $\zeta$ direction converts the circular Dirac cone to rotated elliptical one. The regular expression of this law is restored straining along $\zeta = 0^{\textrm{o}}$ or $90^{\textrm{o}}$, where non-rotated elliptical Dirac cones are obtained. On the other hand, the Snell's law of pseudo-spin  

\begin{equation}
s\rho\sin(\phi - \alpha_x) = s'\rho'\sin(\phi' - \alpha_x),
\label{phi_Snell}
\end{equation}

\noindent is also reduced to the conventional form when $\alpha_x = 0$. An important consequence of this novel refraction law reveals that the conservation of pseudo-spin occurs for $\phi = \phi' = \alpha_x$ regardless the Fermi level. Since the group velocity and pseudo-spin of electrons have different direction in anisotropic systems, perfect transmission emerges in a prefered direction which is not necessarily the normal incidence. In order to confirm this fact the Fresnel-like coefficient is derived using the wavefunction

\begin{equation}
|\Psi_I\rangle = \frac{1}{\sqrt{2}}\begin{pmatrix}
1\\
s\textrm{e}^{i\phi}
\end{pmatrix}\textrm{e}^{ixp_x/\hbar} + 
\frac{1}{\sqrt{2}}r\begin{pmatrix}
1\\
-s\textrm{e}^{-i\overline{\phi}}
\end{pmatrix}\textrm{e}^{-ix\overline{p}_x/\hbar}
\label{pwr1}
\end{equation}   

\noindent in the region I, where the coefficient $r$ is the probability amplitude for the reflected beam. The state $|\Psi_I\rangle$ takes into account the novel refraction law of pseudo-spin \eqref{phirefl}. While in the region II, the transmitted wavefunction is given by

\begin{equation}
|\Psi_{II}\rangle = \frac{1}{\sqrt{2}}t\begin{pmatrix}
1\\
s'\textrm{e}^{i\phi'}
\end{pmatrix}\textrm{e}^{ixp'_x/\hbar},
\label{pwr2}
\end{equation}

\noindent where the amplitude for the transmitted beam is denoted by $t$. The probability amplitudes are calculated from the continuity condition for the wavefunctions \eqref{pwr1} and \eqref{pwr2} at $x = 0$. Solving the equation system for $r$ and $t$ and substituting the relations \eqref{phis}, the transmission probability 

\begin{equation}
T_{pn}(\chi,\chi') = \frac{\cos\chi\cos\chi'}{\cos^2\left[\frac{1}{2}(s'\chi' + s\chi)\right]}
\label{R}
\end{equation}

\noindent straightforwardly confirms that $\phi = \phi'$ leads to $T(0) = 1$. In order to show that the KT of anisotropic massless Dirac fermions in $p$-$n$ homojunctions of USG is angularly shifted when $\zeta \neq 0^{\textrm{o}}$ and $90^{\textrm{o}}$, the Fresnel-like coefficient of electrons $T(\chi,\chi')$ must be expressed as a function of $\theta$ and $\theta'$. Thus, the calculation of group velocity serves for relating the angle of incidence $\theta$ with $\chi$ (see appendix B)

\begin{equation}
\tan\theta = \pm\frac{w_y\cos[\chi \mp s(\alpha_x + \alpha_y)]}{w_x\cos\chi}.
\label{thchi}
\end{equation}

\noindent Again the sign minus is for the angle of reflection $\overline{\theta}$. The relation between angle of refraction $\theta'$ and $\chi'$ can be obtained choosing the sign plus and performing the substitutions $\theta \rightarrow \theta'$, $\chi \rightarrow \chi'$, and $s \rightarrow s'$. Setting $\chi = 0$ in Eq. \eqref{thchi}, the angular shifting of KT 

\begin{equation}
\theta_{KT} = \arctan\left(\frac{w_y}{w_x}\cos(\alpha_x + \alpha_y)\right)
\end{equation}

\noindent is found. Other way to write the reflection and refraction law of electrons in $p$-$n$ homojunctions of USG is given by

\begin{equation}
\tan\overline{\theta} = \tan\theta - 2\tan\theta_{KT}
\label{eoptlr}
\end{equation}
\begin{equation}
\tan\theta' = ss'\tan\theta + (1 - ss')\tan\theta_{KT},
\label{eoptlrr}
\end{equation}

\noindent where the last expression for the refraction law is obtained from the geometrical condition $\rho = \rho'$, which corresponds to the special case of elliptical energy contours with the same vertical half-width at the Fermi level (see appendix B). It is interesting to note that electrons impinging under normal incidence have nonzero angles of reflection and refraction. Since elliptical Dirac cone are rotated in both sides of the junction the outcoming electron beam has a $v_y$ component different to zero, as shown in Figs. \ref{EOD}(b) and \ref{CAVLs}(b). The scattering of electrons in $p$-$n$ homojunctions of USG under the focusing condition $\rho = \rho'$ and $s = -s'$ shall be discussed in the following section. 

\section{Electron tunneling in asymmetric Veselago lenses} 
\label{AVLs}

The $p$-$n$ homojunction of USG satisfying the focusing condition leads to the appearance of an AVL, as shown in Fig. \ref{CAVLs}(a). If tensile strain is applied along $\zeta \neq 0^{\textrm{o}}$ and $90^{\textrm{o}}$ direction, the Dirac cones rotate. This rotation makes asymmetric the converged electron flow. Therefore, the focus is moved out the normal axis [see Fig. \ref{CAVLs} (a)]. Thus, a divergent flow emitted at $(-x_0,0)$ in the region I is focused at the spot $(x_0,y'_0)$. The geometrical relation between $\theta$ and $\theta'$ of AVLs can be obtained from the ray equations $y_I = (x + x_0)\tan\theta$ and $y_{II} = x\tan\theta' + x_0\tan\theta$ for the regions I and II, respectively. The specific relationship $\tan\theta' = -\tfrac{x_0}{x'_0}\tan\theta + \tfrac{y'_0}{x_0}$ is identical to the Snell's law in Eq. \eqref{eoptlrr}. In this way, the $y$ position of focus $y'_0 = x_0\tan\theta_{KT}$ is a function of the strain parameters. 

It is worth to note that the perfect tunneling occurs when electrons fly on the straight line which links from the point source to focus. Although the pseudo-spin and group velocity have different directions, the angles $\phi = \alpha_x$ and $\theta = \theta_{KT}$ are remained in the KT, as shown in Figs. \ref{CAVLs}(a), (b), and (d). Moreover, one part of the incoming particle flow has a positive refraction within incidence range $0 \leq |\theta| \leq |\theta_{KT}|$ [see Figs. \ref{CAVLs}(a), (b), and (c)]. Such a positive refraction is caused by the rotation of the Dirac cones. This atypical result contrasts with the total negative refraction of conventional Veselago lenses \cite{Ho Lee,Cheianov,Pendry}. Symmetrical case is recovered applying uniaxial strain along the $\zeta = 0^{\textrm{o}}$ or $90^{\textrm{o}}$ direction, as shown in Fig. \ref{CAVLs}(c). Other effect is the obtention of reflected electron beams staying on the same straight line than refracted electrons [see Figs. \ref{CAVLs}(a) and (b)]. From the optical laws \eqref{phirefl}, \eqref{phi_Snell}, \eqref{eoptlr}, and \eqref{eoptlrr} is verifiable that the propagation and pseudo-spin orientation of reflected and refracted electrons are opposite, namely $\phi' = -\overline{\phi}$ and $\theta' = -\overline{\theta}$. 

Since specific strain values in AVLs can improve the particle transmission [see Fig. \ref{CAVLs}(d)], the average of Fresnel-like coefficient \eqref{R} regards to the angle of incidence $\theta$

\begin{widetext}
\begin{equation}
\langle T_{pn} \rangle = \frac{w_y\sin(\alpha_x + \alpha_y)[w_x^3 + w_xw_y^2\cos(2\alpha_x + 2\alpha_y) + w_y(w_y^2 - w_x^2)\sin(\alpha_x + \alpha_y)]}{w_x^4 + w_y^4 + 2w_x^2w_y^2\cos(2\alpha_x + 2\alpha_y)}
\label{Avtr}
\end{equation}
\end{widetext}

\noindent is obtained (see appendix C). Thus, Fig. \ref{ATP} shows the particle transmission efficiency as a function of strain parameters. In the range $-30^{\textrm{o}} \leq \zeta \leq 30^{\textrm{o}}$, the particle tunneling is higher than unstrained case when the tensile strain is increased. Whereas for strain values in the interval $\epsilon > 15 \%$ and $45^{\textrm{o}} < \zeta < 75^{\textrm{o}}$ is observed a decreasing in the particle transmission efficiency. This behavior is due to that the pseudo-spin is rotated by the strain. Thus, a strong change in the pseudo-spin orientation causes the increasing of reflection probability. 

\begin{figure}
\includegraphics[trim = 0mm 0mm 0mm 0mm, scale= 0.23, clip]{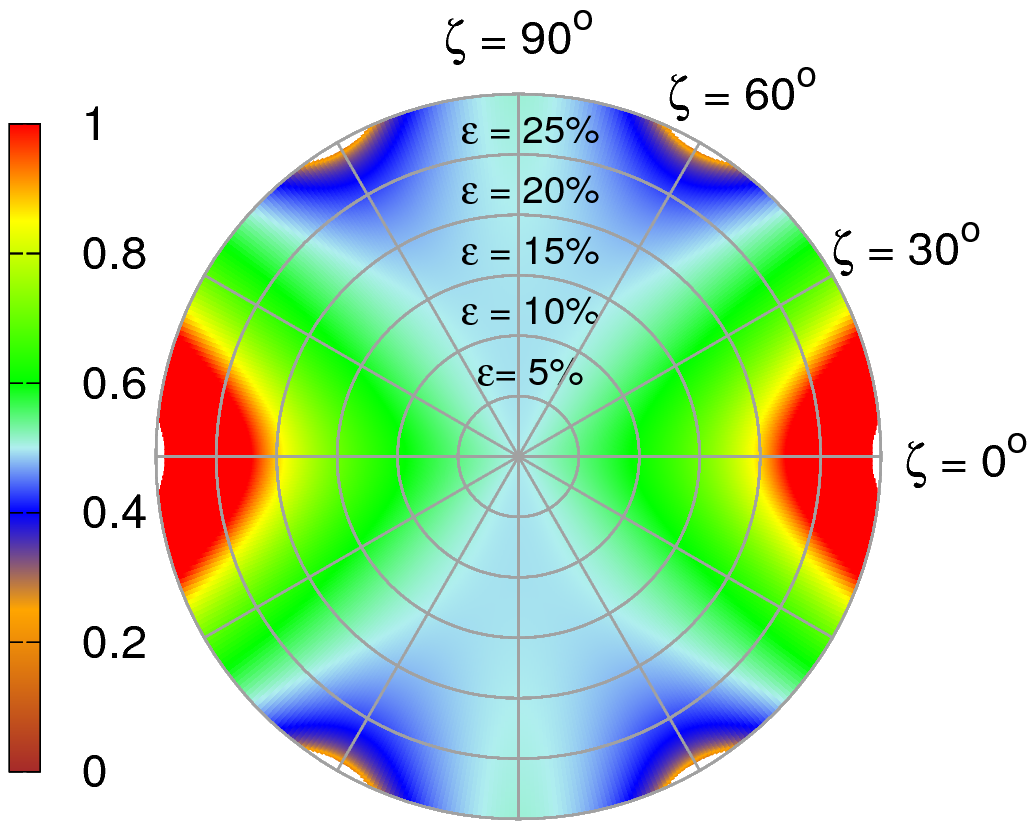} 
\caption{Average transmission probability of the asymmetric Veselago lens as a function of uniaxial strain parameters $\epsilon$ and $\zeta$. High transmission efficiency is found at the range $15 \% \leq \epsilon \leq 29 \%$ and $-15^{\textrm{o}} \leq \zeta \leq 15^{\textrm{o}}$. In contrast, low transmission probability occurs in the same range of $\epsilon$ and angular sector $45^{\textrm{o}} \leq \zeta \leq 75^{\textrm{o}}$.}
\label{ATP}
\end{figure}

\section{Fabry-P\'erot fringes in homojunctions of uniaxially strained graphene}
\label{npn_junc}

\begin{figure*}
\begin{tabular}{ccc}
(a) \qquad \qquad \qquad \qquad \qquad \qquad \qquad \qquad \qquad \qquad \qquad & (b) \qquad \qquad \qquad \qquad \qquad \qquad \qquad \qquad \qquad & (c) \qquad \qquad \qquad \qquad \qquad \qquad \qquad \qquad \qquad \\
\includegraphics[trim = 0mm 0mm 0mm 0mm, scale= 0.32, clip]{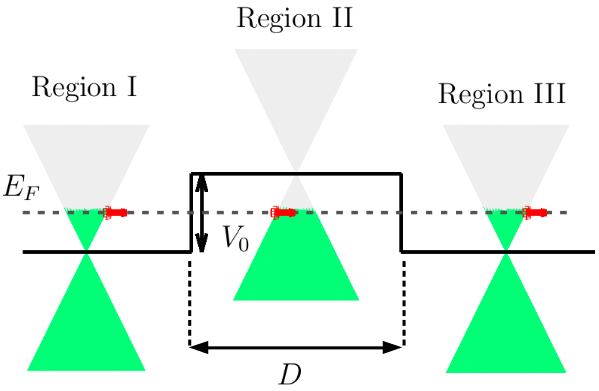} &
\includegraphics[trim = 0mm 0mm 0mm 0mm, scale= 0.14, clip]{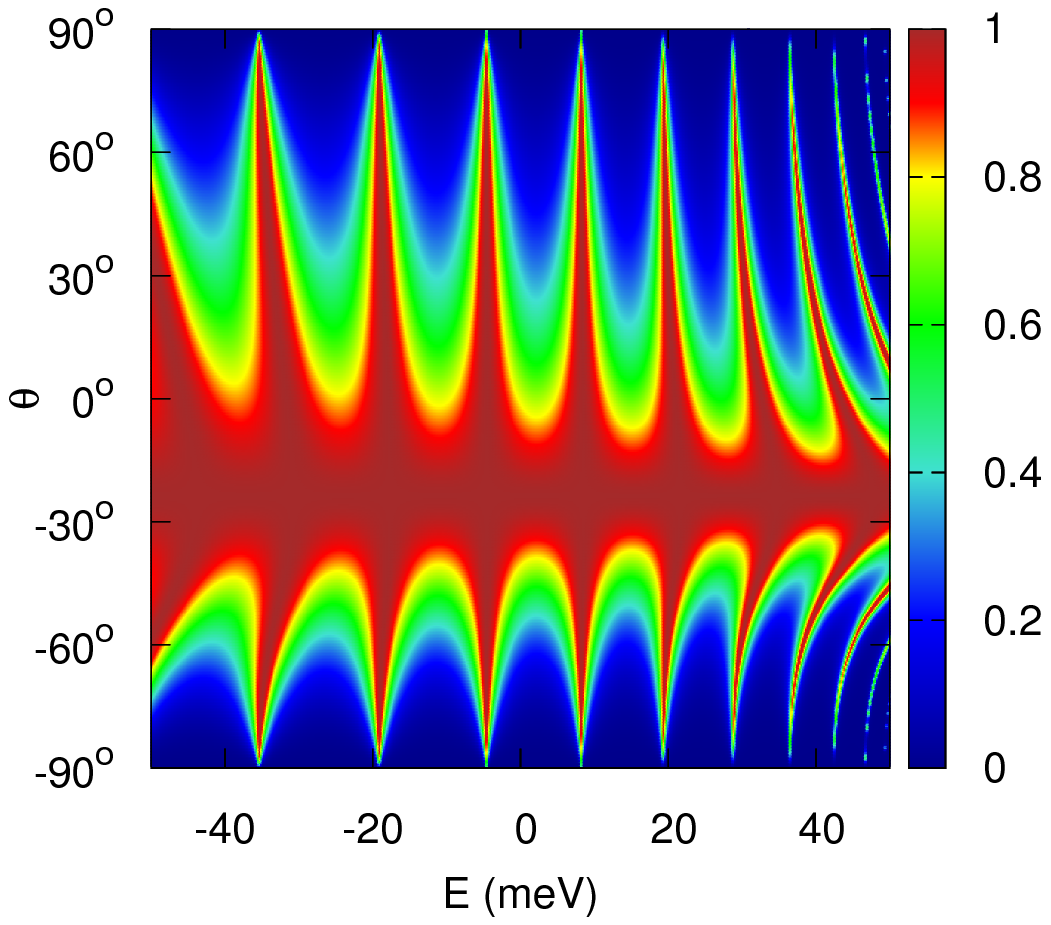}&
\includegraphics[trim = 0mm 0mm 0mm 0mm, scale= 0.14, clip]{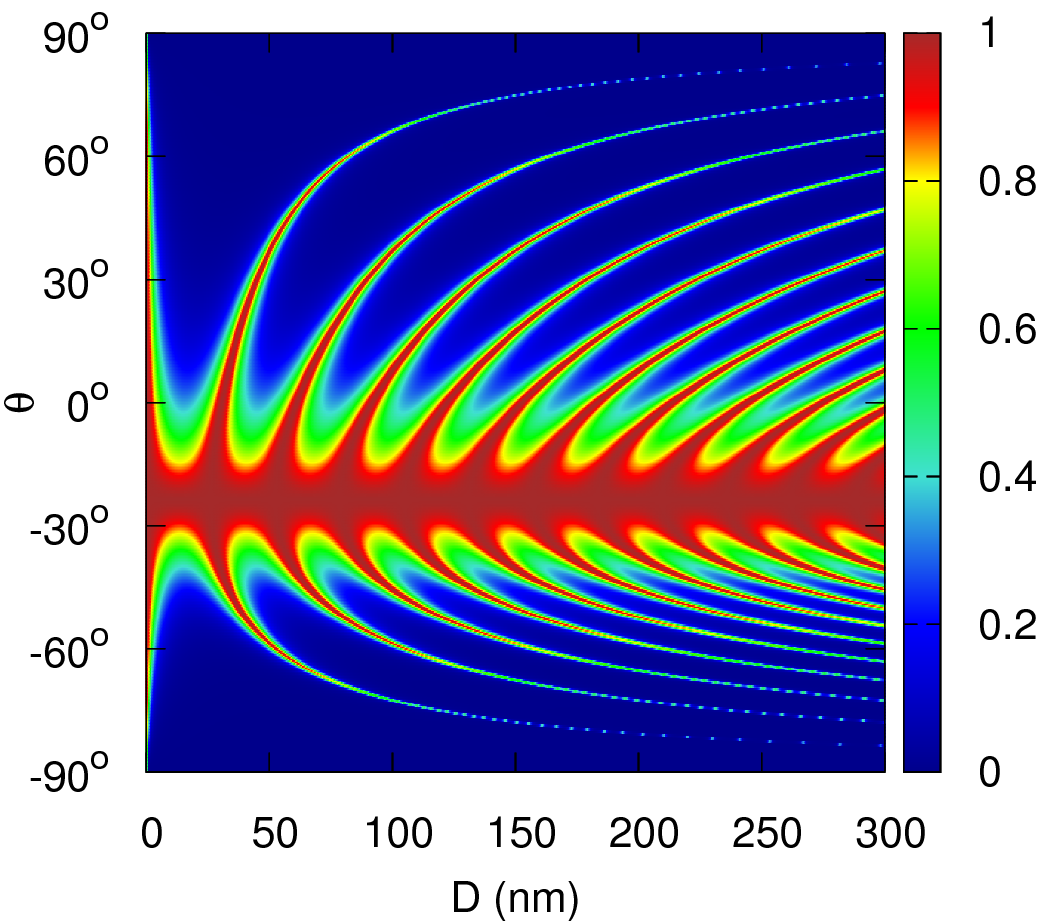}
\end{tabular}
\caption{Description of the particle scattering through an $n$-$p$-$n$ homojunction of uniaxially strained graphene. (a) Schematic representation of the potential barrier and Dirac cone structure. Solid area represents the occupied states and the horizontal dashed line shows the Fermi level. (b) and (c) Transmission probability as a function of $\theta$, $E$, and barrier width $D$ for uniaxially strain values $\epsilon = 23 \%$ and $\zeta = 45^{\textrm{o}}$ and barrier height $V_0 = 100$ meV. The shifted Klein tunneling at $\theta_{KT} = -23.7^{\textrm{o}}$ remains unaffected against variations of $E$ and $D$.}
\label{Barrier}
\end{figure*}

The scattering of anisotropic massless Dirac fermions in $n$-$p$-$n$ homojunctions of USG is considered [see Fig. \ref{Barrier}(a)]. In this system, the regions I and III have the same negative doping level and the region II of width $D$ is positively doped. The coherence length and mean free path must be larger than the device dimensions for guaranteeing ballistic transport \cite{Low2,Cheianov2,Fuchs}. Since the three regions are identically deformed, the Dirac cones have the same position at the reciprocal space. This feature leads to a valley-independent refraction. In order to address the particle scattering problem the same wavefunction \eqref{pwr1} is used. For the region III,  the wavefunction \eqref{pwr2} is identically written deleting the prime on the parameters. While the wavefunction in the region II is given by

\begin{equation}
|\Psi_{II}\rangle = \frac{1}{\sqrt{2}}t'\begin{pmatrix}
1\\
s'\textrm{e}^{i\phi'}
\end{pmatrix}\textrm{e}^{ixp'_x/\hbar} +    
\frac{1}{\sqrt{2}}r'\begin{pmatrix}
1\\
-s'\textrm{e}^{-i\overline{\phi}'}
\end{pmatrix}\textrm{e}^{-ix\overline{p}'_x/\hbar},
\label{pwr3}
\end{equation}   

\noindent where $r'$ and $t'$ are the amplittudes for the wavefunction within the potential barrier. Applying the matching conditions at $x = 0$ and $x = D$ and taking into account the novel electron optics laws of pseudo-spin \eqref{phirefl} and \eqref{phi_Snell}, the transmission probability is 
\begin{eqnarray}
T_{npn}(\chi,\chi') = \qquad \qquad \qquad \qquad \qquad \qquad \qquad \qquad & & \nonumber \\
\frac{\cos^2\chi\cos^2\chi'}{\cos^2\chi\cos^2\chi'\cos^2\overline{\beta} + (1 - ss'\sin\chi\sin\chi')^2\sin^2\overline{\beta}}, & &
\label{T_KT}
\end{eqnarray}

\noindent where $\overline{\beta} = (w_y\rho'D/\hbar w_x)\cos\chi'$ shows the dependency of resonant condition $\overline{\beta} = n\pi$ with the strain. This allow to exhibit non-negligible changes in the interference pattern regards to unstrained graphene transistors. In Figs. \ref{Barrier}(b) and (c), the intraband ($s = s' = -1$) and interband ($s = -s' = 1$) tunneling regime match perfectly due to the gapless band structure. This feature in the particle transmission density is due to the geometry of Dirac cones regardless of the pseudo-spin nature \cite{Betancur2,Betancur4}. In this system, uniaxial tension along the $\zeta \neq 0^{\textrm{o}}$ or $90^{\textrm{o}}$ angularly shifts the KT and breaks the mirror symmetry of Fabry-P\'erot resonances, as shown in Figs. \ref{Barrier}(b) and (c). It is possible to observe that anisotropic massless Dirac fermions have non-resonant tunneling when the angle of incidence is $\theta = \theta_{KT}$. In contrast, electrons impinging the barrier under normal incidence possess resonant tunneling as well as nonzero backscattering probability.

On the other hand, the increasing of tensile strain makes transparent the barrier for tension angles $\zeta$ near the normal direction. This is because the pseudo-spin has a slowly variation in $\theta$. Thus, the conservation of pseudo-spin is slightly affected doing that the deviated KT persists for a wide incidence sector. The contrary case is observed for tension angles far away the normal axis. Again, electrons with $\theta = \theta_{KT}$ perfectly tunnels without deflecting their trajectory and regardless the tunneling regime. However, the high angular variation rate of pseudo-spin reduces the angular robustness of perfect transmission. Such observations in $n$-$p$-$n$ homojunctions are similar to the behavior of average transmission efficiency of AVLs through the tuning of strain parameters in Fig. \ref{ATP}. 

\section{Conclusions and final remarks}
\label{conclusions}
In summary, partial positive refraction in asymmetrical Veselago lenses have been shown in uniaxially strained graphene homojunctions. The effective TB Weyl-like Hamiltonian was put forward for studying the scattering of anisotropic massless Dirac fermions. These particles obey atypical Fresnel-like coefficient, reflection and refraction laws which lead to a novel optical-like phenomena. Thus, electron rays under normal incidence have reflected and refracted beams different to zero. For uniaxial strains breaking the mirror symmetry regards to the normal direction, Klein tunneling is angularly deviated. When the focusing condition is satisfied, the trajectory of perfect transmission occurs in the straight line that connects the point source and focus. Lifting this condition through the Fermi level modulation, the conservation of pseudo-spin warranties the appearance of Klein tunneling. This effect persists in an $n$-$p$-$n$ homojunction varying the barrier's width and doping level. Fabry-P\'erot interference shows asymmetrical resonant tunneling. In contrast to isotropic graphene, electrons under normal incidence can be backscattered. Further, uniaxial strain near the normal axis improves the particle transmission efficiency. This result could be used for avoiding losses by draining in electron optics devices. Therefore, uniaxially strained graphene and related materials are good candidates as platform for the realization of elliptical Dirac optics. The decrease of transmission efficiency using high tensile strains and tension angles far away the normal axis could serve for the implementation of quantum confinement. Asymmetric Veselago lenses may be useful for the design of enhanced scanning tunneling microscopes with strain-bendable probing tips. 

\section*{Acknowledgments}

Y.B.-O. gratefully acknowledges financial support from CONACYT Proyecto Fronteras 952 Transporte en sistemas peque\~nos, cl\'asicos y cu\'anticos. The author also thanks to T. Stegmann, F. Leyvraz, T.H. Seligman, G. Cordourier-Maruri, and R. de Coss for helpful discussions, comments, and critical reading of the manuscript.

\appendix
\setcounter{section}{1}
\section*{Appendix A: Complex velocities in terms of lattice vectors and hopping parameters}
\label{ws}
In this appendix is shown the derivation of complex velocities in terms of lattice vectors and hopping parameters, which are obtained through a TB calculation to nearest neighbors. Using the definition of complex velocities in Eq. \eqref{w} and nearest neighbors positions in terms of lattice vectors $\vec{a}_1$ and $\vec{a}_2$ is found that 

\begin{equation}
w^c_x = \frac{i}{\hbar}\left(a_{1x}t_1\textrm{e}^{-i\vec{K}_D\cdot\vec{\delta}_1} + a_{2x}t_2\textrm{e}^{-i\vec{K}_D\cdot\vec{\delta}_2}\right),
\end{equation}

\noindent where $w^c_y$ has identical expression replacing $x \rightarrow y$. Taking into account the relation 

\begin{equation}
\cos[\vec{K}_D\cdot(\vec{\delta}_1 - \vec{\delta}_2)] = \frac{t^2_3 - t^2_2 - t^2_1}{2t_1t_2},
\end{equation}

\noindent which is obtained from the Dirac points equation $\sum^3_jt_j\textrm{e}^{-i\vec{K}_D\cdot\vec{\delta}_j} = 0$, is possible to prove that square module of $w^c_x$ given by

\begin{equation}
w_x^2 = \frac{1}{\hbar^2}[a^2_{1x}t^2_1 + a^2_{2x}t^2_2 + a_{1x}a_{2x}(t^2_3 - t^2_1 - t^2_2)]
\end{equation}

\noindent can be related with the strain parameters. Calculating $\textrm{Im}\{w^{c*}_xw^c_y\}$, where the operations $\textrm{Im\{\}}$ and * are the imaginary part and complex conjugate respectively, a useful identity of complex velocities and their phases 

\begin{eqnarray}
w_x^2w_y^2\sin^2(\alpha_x + \alpha_y) = \frac{1}{4\hbar^4}|\vec{a}_1\times\vec{a}_2|^2(t_1 + t_2 + t_3) \times &  &\nonumber \\
(-t_1 + t_2 + t_3)(t_1 - t_2 + t_3)(t_1 + t_2 - t_3) \qquad &  &
\label{xi2}
\end{eqnarray} 

\noindent is obtained. Such expressions serve for establishing the brigde between complex velocities and uniaxial strain parameters from the relations \eqref{lattvs}, \eqref{d}, and exponential decay. In this way, the electron optics behavior in homojunctions of uniaxially strained graphene can be described.

\appendix
\setcounter{section}{2}
\section*{Appendix B: Derivation of electron optical laws in rotated Dirac cone systems}

In order to obtain the electron optical laws in systems presenting rotated Dirac cones in their band structure, the conservation of $E$, $p_y$, and $j_x$ must be applied. The definition of pseudo-spin angle in the Weyl-like Hamiltonian \eqref{HD} can be written in terms of linear momentum

\begin{equation}
\tan\phi = \frac{w_xp_x\sin\alpha_x + w_yp_y\sin\alpha_y}{w_xp_x\cos\alpha_x - w_yp_y\cos\alpha_y}.
\label{defphi}
\end{equation}

\noindent Using the dispersion relation 
\begin{equation}
|E - V| = \sqrt{w_x^2p^2_x + w_y^2p^2_y + 2w_xw_yp_xp_y\cos(\alpha_x + \alpha_y)}
\label{ldr}
\end{equation}

\noindent and the effective Snell's law $p_y = \rho\sin\chi = \rho'\sin\chi'$, the $x$ component in the linear momentum is given by

\begin{equation}
p_x = \pm s\frac{w_y}{w_x}\rho\sin(\alpha_x + \alpha_y \mp \chi),
\end{equation} 

\noindent where $\rho = |E - V|/w_y(\sin(\alpha_x + \alpha_y))$ is the effective refraction index. Substituting the linear momentum components in Eq. \eqref{defphi}, the relations of $\phi$ in Eq. \eqref{phis} as a function of $\chi$ are found. 

The optical laws of electron propagation are derived determining the group velocity

\begin{equation}
v_x =  \partial_{p_x}E = \frac{1}{s|E - V|}[w_x^2p_x + w_xw_yp_y\cos(\alpha_x + \alpha_y)],
\end{equation}

\noindent where the dispersion relation \eqref{ldr} of effective Weyl-like Hamiltonian \eqref{HD} is considered. An identical expression for $v_y$ is found developing $v_y = \partial_{p_y}E$. The direction of electron beam in Eq. \eqref{thchi} is calculated, performing the ratio $v_y/v_x$ and substituting the components of linear momentum in terms of $\chi$ for the cases of incidence, reflection, and refraction. It is always possible to delete the dependency of $\chi$ in the reflection law \eqref{eoptlr} when the angles $\theta$ and $\overline{\theta}$ are related. In the electron refraction law for the AVLs \eqref{eoptlrr}, the application of focusing condition removes the $\chi$-dependency. 

\appendix
\setcounter{section}{3}
\section*{Appendix C: Average Fresnel-like coefficient in asymmetric Veselago lenses}

Since the transmission probability in AVLs is $T_{pn}(\chi) = \cos^2\chi$, the average of this quantity regards to angle of incidence $\theta$ is written as

\begin{equation}
\langle T_{pn} \rangle = \frac{1}{\pi}\int^{\frac{\pi}{2}}_{-\frac{\pi}{2}} \cos^2\chi d\theta.
\label{ave}
\end{equation}

\noindent For integrating in $\chi$ is necessary to use 

\begin{equation}
\frac{d\theta}{d\chi} = \frac{w_xw_y\sin(\alpha_x + \alpha_y)}{w_x^2\cos^2\chi + w_y^2\cos^2(\chi - \alpha_x - \alpha_y)},
\end{equation}

\noindent which is derived from the expression \eqref{thchi}. Using the following defined integral \cite{Gradstheyn}

\begin{eqnarray}
\int^{\frac{\pi}{2}}_{-\frac{\pi}{2}} \frac{\cos^2\chi d\chi}{a\cos^2\chi + 2b\sin\chi\cos\chi + c\sin^2\chi} = & & \nonumber\\
\frac{\pi}{4b^2 + (a - c)^2}\left[a - c + \frac{2b^2 - (a - c)c}{\sqrt{ac - b^2}}\right], & &
\end{eqnarray}

\noindent with parameters $a = w_x^2 + w_y^2\cos^2(\alpha_x + \alpha_y)$, $b = w_y^2\sin(\alpha_x + \alpha_y)\cos(\alpha_x + \alpha_y)$, and $c = w_y^2\sin^2(\alpha_x + \alpha_y)$, the expression of average Fresnel-like coefficient in Eq. \eqref{Avtr} is obtained substituting the integral value in Eq. \eqref{ave}.

\end{document}